\documentclass[final,5p,times]{elsarticletc}

\usepackage{amsmath,amssymb,amsfonts,amscd}
\usepackage{graphicx}

\usepackage{color,soul}

\begin{document}

\begin{frontmatter}

\title{Exotic quantum holonomy in Hamiltonian systems}
\author[label1]{Taksu Cheon \corref{cor1}}
\ead{taksu.cheon@kochi-tech.ac.jp}
\author[label2]{Atushi Tanaka}
\ead{tanaka-atushi@tmu.ac.jp}
\author[label3]{Sang Wook Kim}
\ead{swkim0412@pusan.ac.kr}
\address[label1]{Laboratory of Physics, Kochi University of Technology
Tosa Yamada, Kochi 782-8502, Japan}
\address[label2]{Department of Physics, Tokyo Metropolitan University,
Hachioji, Tokyo 192-0397, Japan}
\address[label3]{Department of Physics Education, Pusan National University,
Busan 609-735, South Korea}
\cortext[cor1]{corresponding author}

\date{September 11, 2009}

\begin{abstract}
We study the evolution of quantum eigenstates in the presence of level crossing 
under adiabatic cyclic change of environmental parameters. 
We find that exotic holonomies, indicated by exchange of the eigenstates after 
a single cyclic evolution, can arise from non-Abelian gauge potentials among 
non-degenerate levels.  
We illustrate our arguments with solvable two and three level models.
\end{abstract}


\end{frontmatter}


\section{Introduction}

Berry phase phenomena are known to be wide-spread, intriguing, and useful in
controlling quantum systems \cite{BE84,SW89}.
The Wilczek-Zee variation,
in which different eigenstates sharing a degeneracy are turned into each other
after the cyclic variation of environmental parameter \cite{WZ84},
is found particularly useful, since it supplies the basis for so-called
holonomic quantum computing \cite{ZR99}.
Whether such state transformations require the existence
of degeneracy throughout the parameter variation,
is a matter in need of further analysis, although it has been the wide-spread assumption.

The exotic holonomies are defined as exchange of eigenstates after one 
period of cyclic parametric evolution without any relevant degeneracy. 
They have been found in time-periodic systems \cite{TM07, MT07} and in
singular systems \cite{CH98,TF01}, but not in finite Hamiltonian systems up to now. 
One obvious reason is that the two real numbers in a real axis cannot be continuously 
exchanged without colliding with each other, i.e. degeneracy. 
Note that in a time periodic system described by a unitary matrix, 
its eigenvalues are complex number on the 
unit circle in the complex plane so that the eigenvalues smoothly exchange their 
position without crossing over each other.
Once level crossing is allowed during the variation of environmental parameter, even in Hamiltonian system there is a possibility of the eigenlevels being exchanged after
the cyclic parameter variation even among non-degenerate levels, which would be 
instrumental in enlarging the resource for holonomic quantum control.

It appears that this is
related to a myth that adiabatic parameter variation excludes level crossing,
and therefore, there is no possibility for exotic quantum holonomies
for non-singular Hamiltonian system.
An argument often raised against systems with level crossing is that
it is not generic, and it represents a set of measure zero in parameter space
of all systems.
However, systems of our interest often lives in a world
with some symmetry, exact or approximate.  Under certain circumstances,
a system can always pass through the point of symmetry
when an environmental parameter is varied along  a circular path.
Then, the level crossing become a common feature rather than an exception.

Surprisingly, it has been known for quite sometime that the adiabatic theorem
is extendable to the case of crossing levels \cite{KA50}.
The eigenstates change smoothly as functions of environmental parameter
even when the level crossing takes place, and this opens up the possibility for eigenvalue 
holonomy for non-singular Hamiltonian systems.
In this article, we explicitly construct such models, which seem to have 
immediate extension to $N$ level cases.
The solvability of our model allows us to examine the analytic
structure of the gauge potentials, which is known to be the mathematical
origin behind the existence of exotic holonomy \cite{KCT09, JS04}.

\section{Adiabatic level crossing and exotic holonomy}

Consider a Hamiltonian system with an environmental parameter,
which we call $\theta$.
We assume, for the moment, that the eigenvalues $E_n(\theta)$ and
eigenstates $\Psi_n(\theta)$ are
non-degenerate for all possible values of the parameter.
\begin{figure}[h]
  \centering
  \includegraphics[width=5cm]{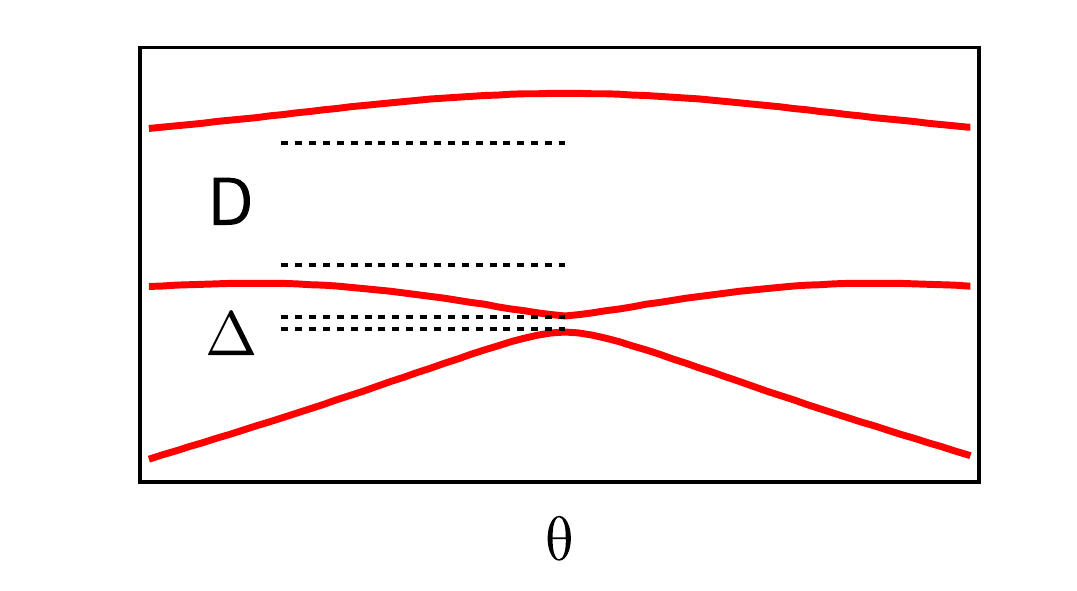}
  \caption{Schematic diagram showing adiabatic level-crossing. $\Delta$ and $D$ represent energy gaps.}
  \label{fig:m1}
\end{figure}
Let us suppose two levels very closely approach each other at a certain value of the parameter,
which can be thought of as an avoided crossing with the closest energy
gap $\Delta$.
%
Let us further assume that the two levels are separated
apart from the next closest level (See Fig. 1) by $D$.
%
%
%
Consider a smooth
cyclic variation of $\theta(t)$ with a period $\tau$, namely, $\theta(\tau)=\theta(0)$. 
Specifically, we require that $\theta(t)$ starts smoothly at the
beginning $t=0$ and stops gently at $t=\tau$ \cite{MO07} to ensure the applicability of the Landau-Zener formula \cite{LA32, ZE32}.
Let us assume that we have inequalities,
\begin{eqnarray}
\frac{1}{D} \ll \tau \ll \frac{1}{\Delta}.
\end{eqnarray}
The period $\tau$ is small enough compared with the inverse of the energy gap $\Delta$
so that the levels completely cross over the gap during the parametric variation,
while it is large enough to ignore any transition among levels except the interacting two levels considered here,
which is the reason why we call this process adiabatic.
The situation remains intact even when the levels do cross according to
some exact symmetry, $\Delta=0$, instead of showing avoided crossing.
In fact, although a tiny avoided crossing caused by a slight symmetry-breaking takes place, the afore mentioned adiabatic level cross-over is robust irrespective of small parametric perturbation.
This is the physics behind the so-called {\em adiabatic  level crossing} \cite{KA50}.
%

Once such an adiabatic level crossing occurs, there is no reason to
assume that each quantum eigenstate should come back to
the corresponding initial state after a cyclic parametric variation.
Only requirement is that the entire set of eigenstates should be the same as before, since the solutions of
eigenvalue equation with a given parameter are uniquely determined. It is thus allowed that the two levels
are exchanged after the parametric variation.
With a parameter $\theta$, which describes the cyclic parametric variation
along the path $C$ from $\theta_i$ to $\theta_f$
satisfying $H(\theta_i) =H(\theta_f)$, such a transition is
described by the holonomy matrix $M$ as \cite{CT09, TC09}
\begin{eqnarray}
\Psi_n(\theta_f) = \sum_m M_{n,m} \Psi_m(\theta_i) e^{-i \phi_m}
\end{eqnarray}
with
\begin{eqnarray}
\label{MHOL}
M = T^* e^{-i\int_C d\theta A(\theta)} \, T e^{i\int_C d\theta A^D(\theta)}
\end{eqnarray}
where $T$ and $T^*$ represents the path-ordering and anti-path ordering of
operator integrals,
the $A(\theta)$ is the non-Abelian gauge potential
\begin{eqnarray}
A_{n,m}(\theta) =
\left< \Psi_n(\theta) \right| i \partial_\theta \!\left. \Psi_m(\theta) \right>,
\label{eq:gauge_potential}
\end{eqnarray}
and $A^D(\theta)$ its diagonal reduction
\begin{eqnarray}
A^D_{n,m}(\theta) = A_{n,n}(\theta) \delta_{n,m}.
\end{eqnarray}
The dynamical phase $\phi_n$ depends on the precise history of the parameter variation, while the holonomy matrix $M$ is solely determined
by the geometry of the path $C$ in the parameter space.
Non-zero off-diagonal component of $M$, if any, signifies the existence of
exotic holonomy.
Physical requirement that an eigenstate does not split with adiabatic parameter variation limits the form of $M$ to be permutation matrix supplemented by possible Manini-Pistolesi off-diagonal phases \cite{MP00, FS05} for each non-zero elements.
Namely, there is only a single non-zero entry to each raw and each column,
and the absolute value of this entry is one.

\section{Two-level model with exotic holonomy}

Consider a two level quantum system described by a parametric Hamiltonian
\begin{eqnarray}
H(\theta) = R(\theta) \left[ \cos\frac{\theta}{2}  Z^{(2)}
+ v \sin\frac{\theta}{2}  F^{(2)} \right]
\label{H_2d}
\end{eqnarray}
with a real number $v$,
\begin{eqnarray}
Z^{(2)} = \sigma_z = \begin{pmatrix} 1 & 0 \cr 0 & -1 \end{pmatrix} ,
\quad
F^{(2)} = I^{(2)}+\sigma_x = \begin{pmatrix} 1 & 1 \cr 1 & 1 \end{pmatrix} ,
\end{eqnarray}
and anti-periodic function $R(\theta)$ with period $2\pi$,
\begin{eqnarray}
R(\theta+2\pi) = -R(\theta).
\end{eqnarray}
A convenient choice we adopt in the numerical examples is $R(\theta)=\cos\frac{\theta}{2}$. 
The system thus becomes $2\pi$ periodic;
\begin{eqnarray}
H(\theta+2\pi) = H(\theta),
\end{eqnarray}
and the parameter $\theta \in [0,2\pi)$ forms a ring, $S^1$.
Note that if (\ref{H_2d}) is written in the form $H(x,y) =x\, Z^{(2)} + y\, v F^{(2)}$, 
the parametric evolution $\theta:0 \to 2\pi$ represents a circle on $(x, y)$ plane 
with its center shifted by the radius into $x$ axis, 
namely $x=1+\cos\theta$ and $y=\sin\theta$, so that it touches the origin. 
If we were to vary $R \in (-\infty, \infty)$ and $\theta \in [0, 2\pi)$ independently, 
the entire plane $(x, y)$ is covered.

The eigenvalue equation
\begin{eqnarray}
H(\theta) \Psi_n(\theta) = E_n(\theta) \Psi_n(\theta)
\end{eqnarray}
is analytically solvable with eigenvalues given by
\begin{eqnarray}
E_n(\theta) = R(\theta) \cos\frac{\theta}{2} P_n(\theta),
\end{eqnarray}
and eigenstates
\begin{eqnarray}
\label{x2wf}
\Psi_n(\theta) =
 \frac{ 1 }{ \sqrt{2 P_n(\theta)^2+2} }
 \begin{pmatrix} P_n(\theta)+1 \cr P_n(\theta)-1 \end{pmatrix},
\end{eqnarray}
for $n = 1, 2$, with
\begin{eqnarray}
P_n(\theta) = v \tan\frac{\theta}{2}
 + (-)^n {\rm sgn}\left[ { v \cos\frac{\theta}{2} } \right]
 \sqrt{ 1+v^2 \tan^2\frac{\theta}{2} } .
\end{eqnarray}
\begin{figure}[h]
  \centering
  \includegraphics[width=5.5cm]{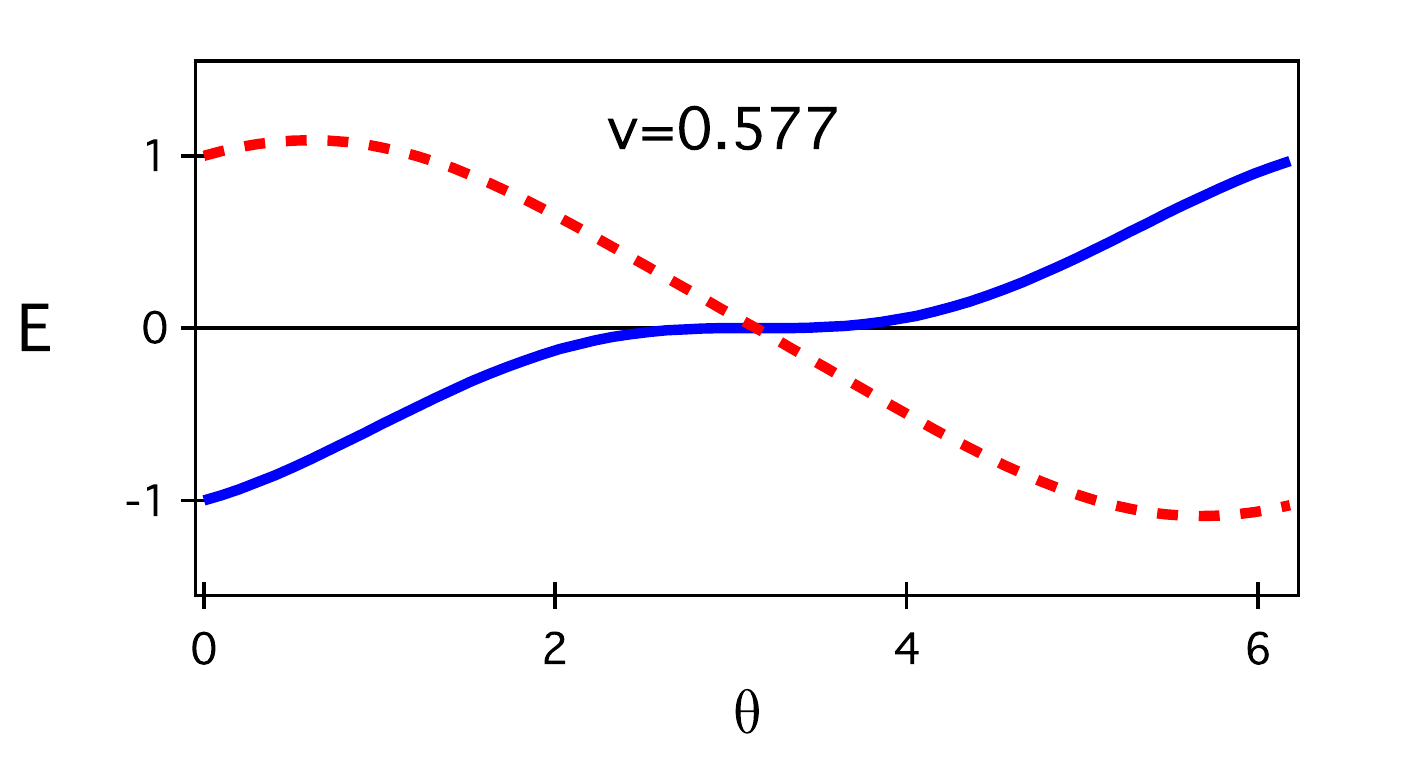}
  \caption{Energy eigenvalues $E_n(\theta)$ of the model (6) 
  as a function of $\theta$ with $v=\frac{1}{\sqrt{3}}$.}
  \label{fig:m2}
\end{figure}

Energy eigenvalues as a function of $\theta$ are shown in Fig.2, in which the most notable
feature is the occurrence of degeneracy at $\theta=\pi$ and the related exchange
of eigenvalues. It guarantees that the set of eigenvalues at $\theta=0$ is
equivalent to that at $\theta=2\pi$, which is a direct
result of $H(0)$ being identical to $H(2\pi)$.
The degeneracy of eigenvalues at $\theta=\pi$ is a direct consequence of
vanishing Hamiltonian, $H(\pi)=0$.
%
%

We note that $P_n(\theta)$ is $4\pi$-periodic. Moreover, we have
%
$P_1(\theta+2\pi) = P_2(\theta)$
and
$P_2(\theta+2\pi) = P_1(\theta)$,
%
which leads to the appearance of exotic holonomy, i.e.,
\begin{eqnarray}
E_1(\theta+2\pi) = E_2(\theta) ,
\quad
E_2(\theta+2\pi) = E_1(\theta) ,
\end{eqnarray}
and also,
$\Psi_1(\theta+2\pi) \propto \Psi_2(\theta)$ and
$\Psi_2(\theta+2\pi) \propto \Psi_1(\theta)$.

The structure of the eigenstates becomes clearer with the re-parameterization
of $P_n(\theta)$ with new angle variable $\chi = \chi(\theta)$, which we define as
\begin{eqnarray}
P_n(\chi) = \tan\frac{\chi+(2n-3)\pi}{4} ,
\end{eqnarray}
namely, $P_1(\chi) = \tan\frac{\chi-\pi}{4}$
and $P_2(\chi) = \tan\frac{\chi+\pi}{4}$.
The monotonously increasing function $\chi(\theta)$ maps
$\theta \in [0, 2\pi)$ to $\chi \in [0, 2\pi)$.
The eigenstates is written, with the new angle parameter $\chi$, as
\begin{eqnarray}
\label{x2wff}
\Psi_n(\chi) =
 \begin{pmatrix} \sin\frac{\chi-(2n-6)\pi}{4} \cr -\cos\frac{\chi-(2n-6)\pi}{4} \end{pmatrix},
\end{eqnarray}
namely
\begin{eqnarray}
\Psi_1(\chi) =
 \begin{pmatrix} -\sin\frac{\chi}{4} \cr \cos\frac{\chi}{4} \end{pmatrix},
\quad
\Psi_2(\chi) =
 \begin{pmatrix} \cos\frac{\chi}{4} \cr \sin\frac{\chi}{4} \end{pmatrix}.
\end{eqnarray}
Note that $\Psi_n(\theta)$ is {\em $8\pi$-periodic},
with anti-periodicity of period $4\pi$.
This is not immediately evident from the expression (\ref{x2wf})
which is in fact discontinuous at several values
of $\theta=n \pi$, and had to be amended with
factor $-{\rm sign}(\sin\frac{\theta-(-)^n\pi}{4})$ to turn into (\ref{x2wff}).
A numerical example of eigenstates as functions of $\theta$ is depicted in Fig. 3.
\begin{figure}[h]
  \centering
  \includegraphics[width=6cm]{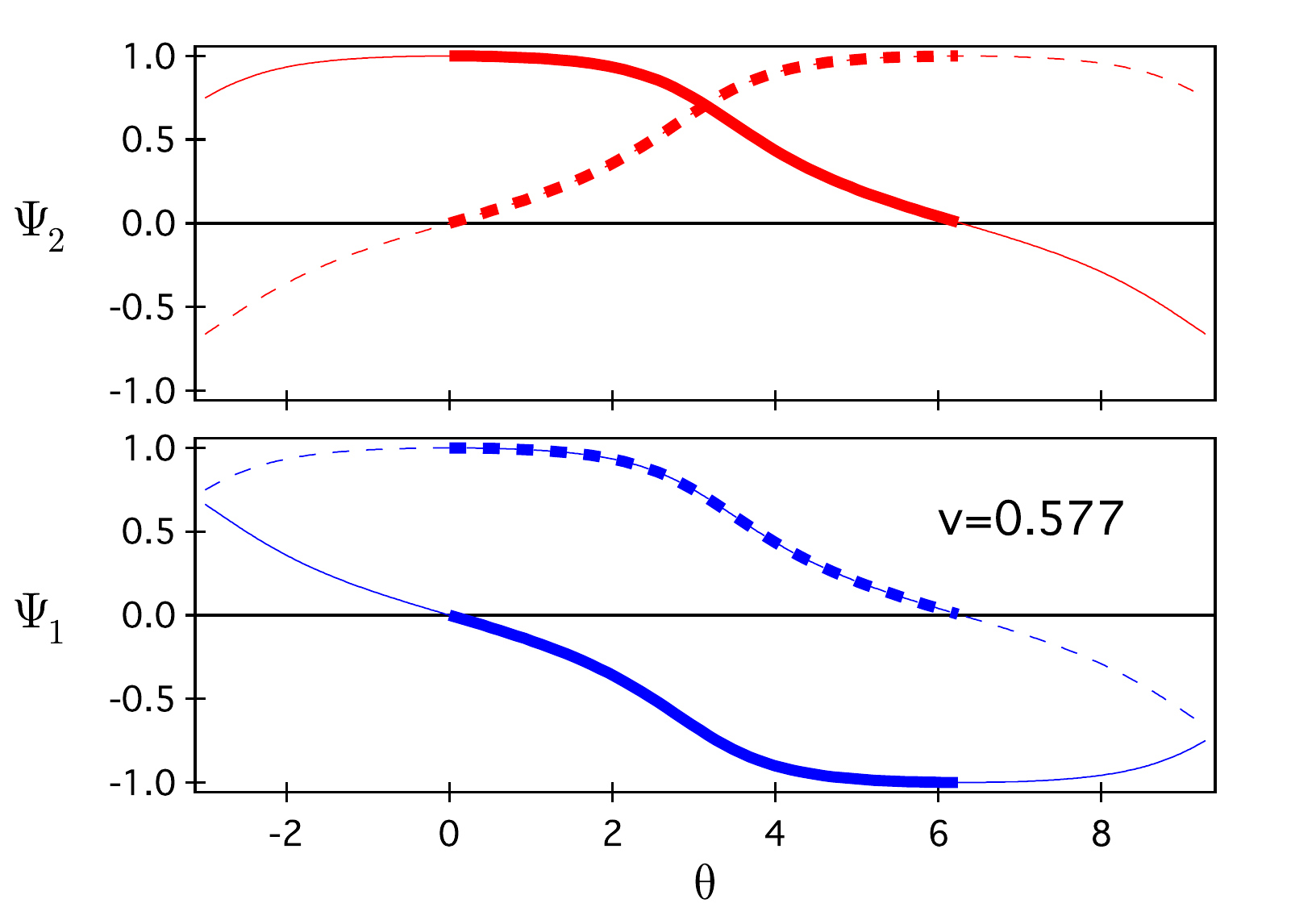}
  \caption{Eigenstates $\Psi_1(\theta)$ (bottom) and $\Psi_2(\theta)$ (top)
   of the model (6) with $R(\theta)=\cos\frac{\theta}{2}$, and $v=\frac{1}{\sqrt{3}}$. 
  The solid and the dashed lines represent the upper and the lower component of 
  the eigenvectors, respectively, both of which are chosen to be real. The range 
  indicated by thick lines represents a single period $\theta \in [0, 2\pi]$.  The values
  outside of this range are shown to display the periodicities
  and mutual relations of $\Psi_1(\theta)$ and
  $\Psi_2(\theta)$}
  \label{fig:m3}
\end{figure}
Adiabatic change of eigenstates are determined by
the gauge potential $A_{nm}(\theta)$  given by
\begin{eqnarray}
A(\theta) =
 \left[ \begin{array}{cc} 0 & -i \cr  i & 0 \end{array}
 \right]  f(\theta)
\end{eqnarray}
where
\begin{eqnarray}
f(\theta) = \frac{1}{4} \frac{\partial \chi(\theta)}{\partial \theta} .
\end{eqnarray}
In Fig. 4, we depicts two examples of function $f(\theta)$.
Obviously, we have $\int_{0}^{2\pi}d\theta f(\theta) = \frac{\pi}{2}$, and we obtain
the holonomy matrix
\begin{eqnarray}
M =
 \left[ \begin{array}{cc} 0 & 1 \cr  -1 & 0 \end{array} \right]
\end{eqnarray}
showing the exotic holonomy with Manini-Pistolesi phase $(-1)^n$.
\begin{figure}[h]
  \centering
  \includegraphics[width=4.2cm]{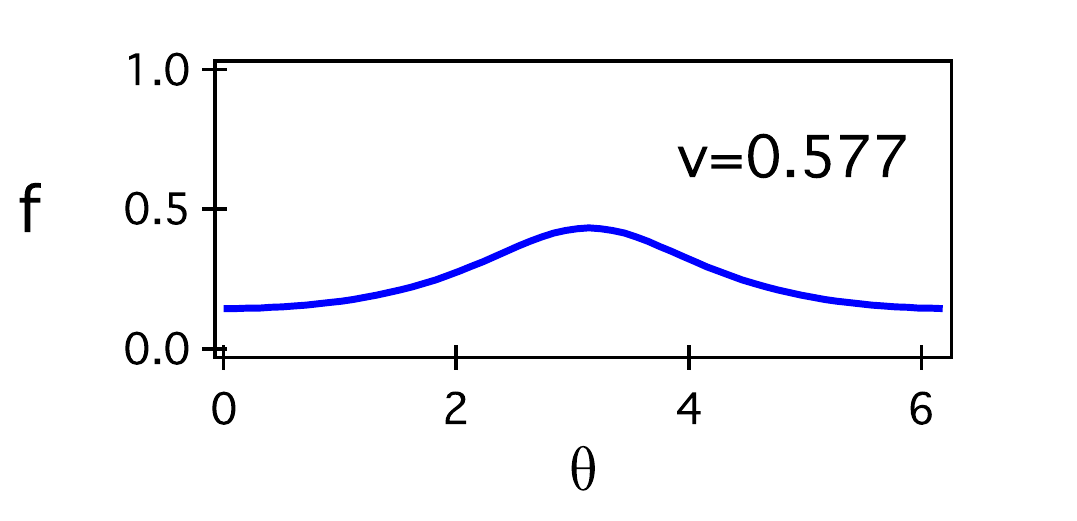}
  \caption{Functional form of gauge potential $f(\theta)$ of the model (6)
  with $v=\frac{1}{\sqrt{3}}$. }
  \label{fig:m4}
\end{figure}
%

%
\begin{figure}[h]
  \centering
  \includegraphics[width=3.5cm]{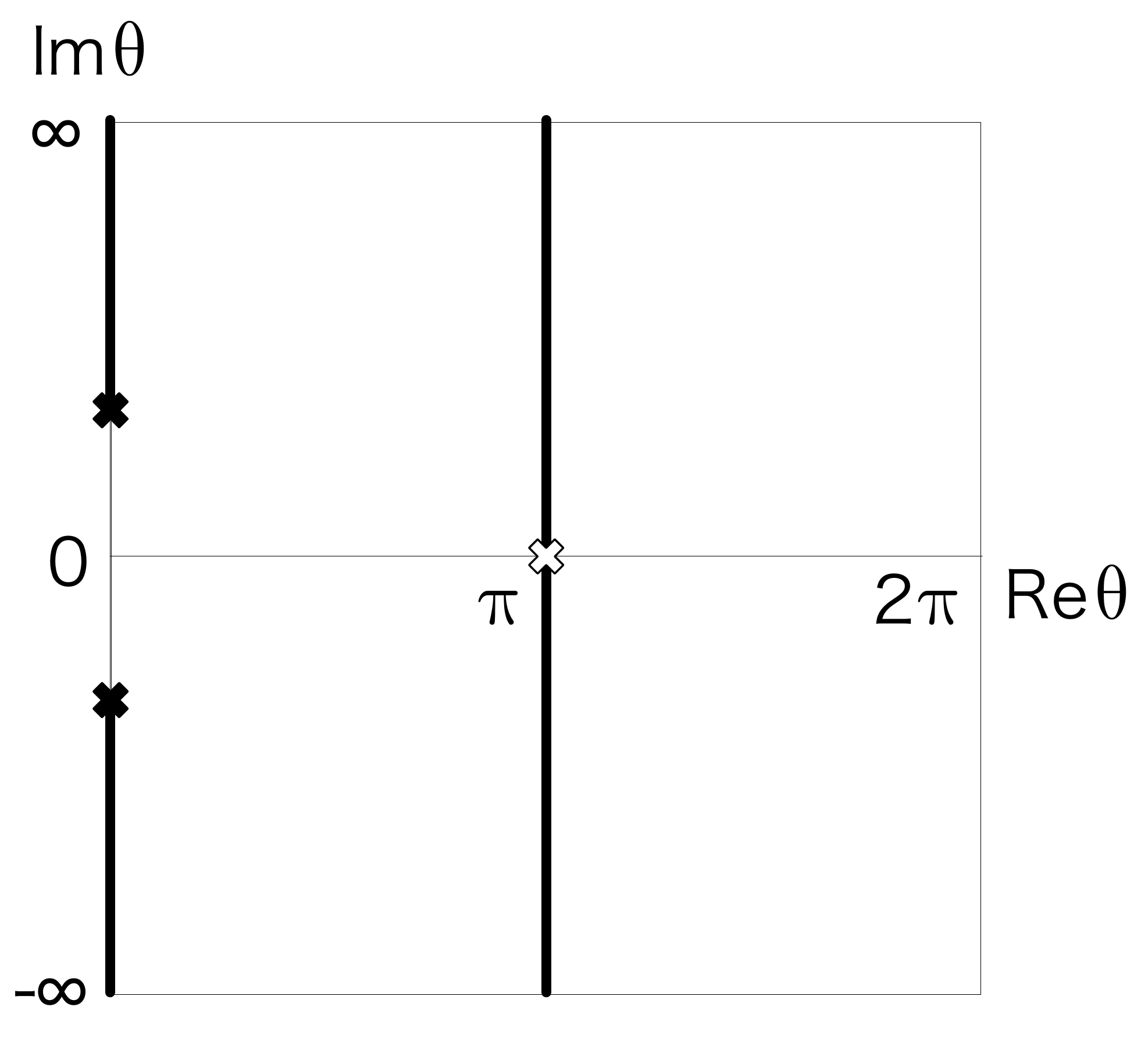}
  \includegraphics[width=3.5cm]{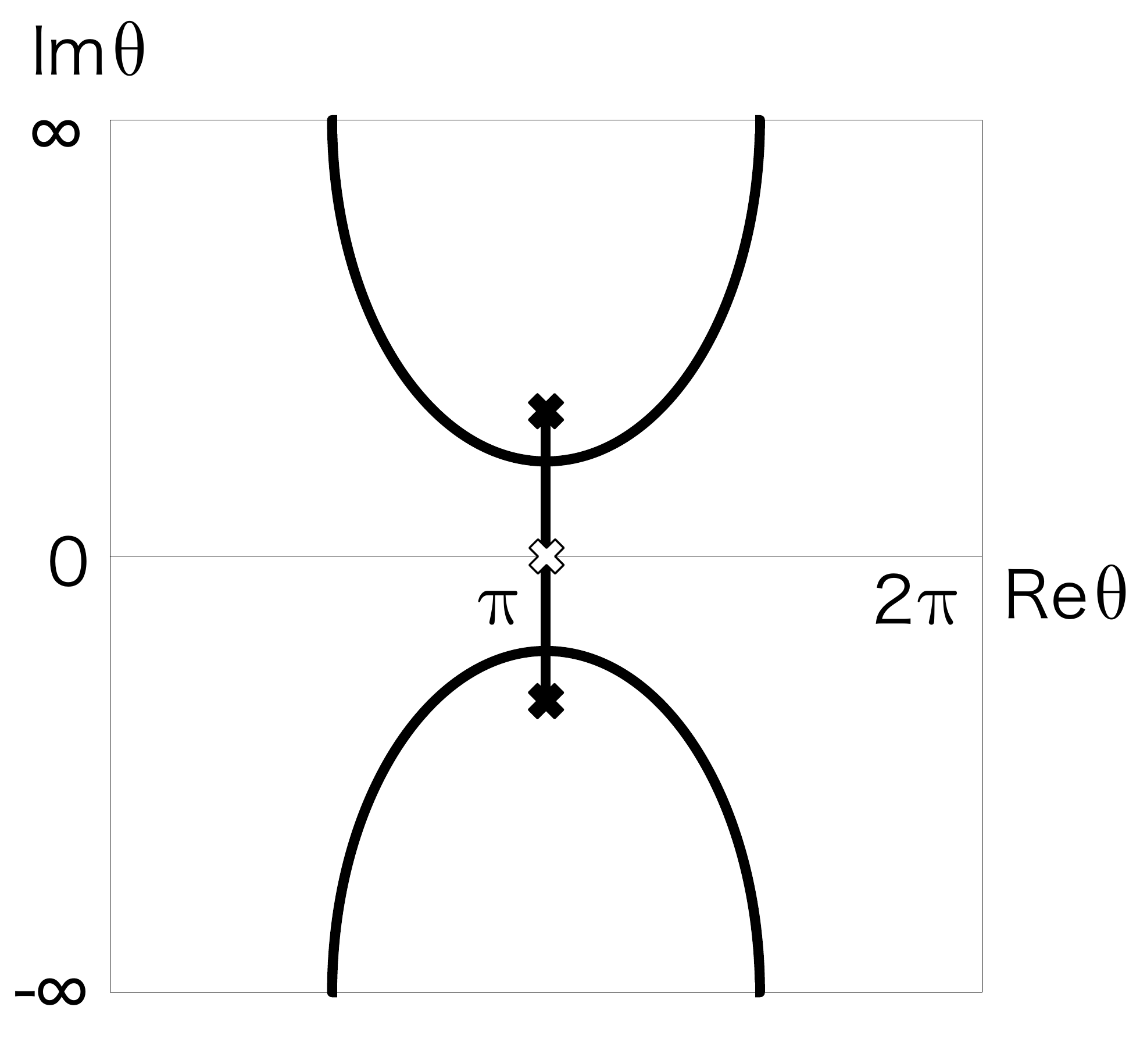}
  \caption{Exceptional points on the Mercator projection of the complex $\theta$ plane
  of system described by (6) with $v>1$ (left) and $v<1$ (right).
  The filled crosses represent the exceptional points that are the poles of gauge
  potential $A(\theta)$, while the unfilled crosses are the points of eigenvalue
  degeneracy which has 
  no effect on the singular behavior of $A(\theta)$.  The solid lines 
  are the branch cuts on which ${\rm Re}(E_2-E_1)=0$.
   In the limit $v \to 1$, the two complex exceptional points (the filled crosses) move 
   to $\pm i\infty$.}
  \label{fig:m5}
\end{figure}
Nontrivial holonomy is known to be related to the analytic structure of
the gauge potential $A(\theta)$ in the complex $\theta$ plain \cite{KCT09},
specifically, its singularities.  These singularities can arise at the
exceptional points $\theta^\star$, which is defined as the point where two complex
energy coincide; $E_m(\theta^\star)-E_n(\theta^\star)=0$.
In our example,
\begin{eqnarray}
R(\theta^\star)\cos\frac{\theta^\star}{2}\left( P_2(\theta^\star)-P_1(\theta^\star)\right)=0
\end{eqnarray}
which yields
\begin{eqnarray}
\theta^\star=\pi,
\quad
\theta^\star_\pm = 2 {\rm arccot} \left(\mp i  v \right),
\end{eqnarray}
where the first one coming from  $R(\theta^\star)\cos\frac{\theta^\star}{2}=0$,
and the $\theta^\star_\pm$ from  $P_2(\theta^\star)-P_1(\theta^\star)=0$.  The exceptional points coming from $R(\theta^\star)\cos\frac{\theta^\star}{2}=0$
has obviously no bearing
on the singularity of $A(\theta)$, while at $\theta^\star_\pm$
%
%
we have the poles of $A(\theta)$ in the form
\begin{eqnarray}
A_{12}(\theta) = -A_{21}(\theta)
\approx \mp \frac{i}{4} \frac{1}{\theta-\theta^\star_\pm}
\qquad
(\theta \to \theta^\star_\pm).
\end{eqnarray}
The existence of the poles explains the non-vanishing values
of $\oint d\theta A_{12}(\theta)$ and $\oint d\theta A_{12}(\theta)$
around the real axis, i.e.
$M_{12}$, $M_{21}\ne 0$ implying the existence of exotic holonomy.
Fig. 5 shows the locations of poles and branching structure of energy
surface $E_n(\theta)$ in Mercator representation of complex
parameter plane $\theta$.

\section{Three-level model with exotic holonomy}

Let us now consider a three level quantum system described by
a parametric Hamiltonian
\begin{eqnarray}
\label{3bdh}
H(\theta) = R(\theta) \left[ \cos\frac{\theta}{2}  Z^{(3)}
+ v \sin\frac{\theta}{2}  F^{(3)} \right]
\end{eqnarray}
where $v$ is real,
\begin{eqnarray}
\label{ZF3}
Z^{(3)} =  \begin{pmatrix} 1 & 0 & 0 \cr 0 & 0 & 0 \cr 0 & 0 & -1 \end{pmatrix} ,
\quad
F^{(3)} =  \begin{pmatrix} 1 & 1 & 1 \cr 1 & 1 & 1 \cr 1 & 1 & 1 \end{pmatrix} ,
\end{eqnarray}
and $R(\theta)$ is anti-periodic with period $2\pi$, i.e.
\begin{eqnarray}
R(\theta+2\pi) = -R(\theta).
\end{eqnarray}
As before, in the numerical examples, we adopt
$R(\theta)=\cos\frac{\theta}{2}$.
The system then becomes $2\pi$ periodic;
\begin{eqnarray}
H(\theta+2\pi) = H(\theta),
\end{eqnarray}
and the parameter $\theta \in [0,2\pi)$ forms a ring, $S^1$.
\begin{figure}[h]
  \centering
  \includegraphics[width=5.5cm]{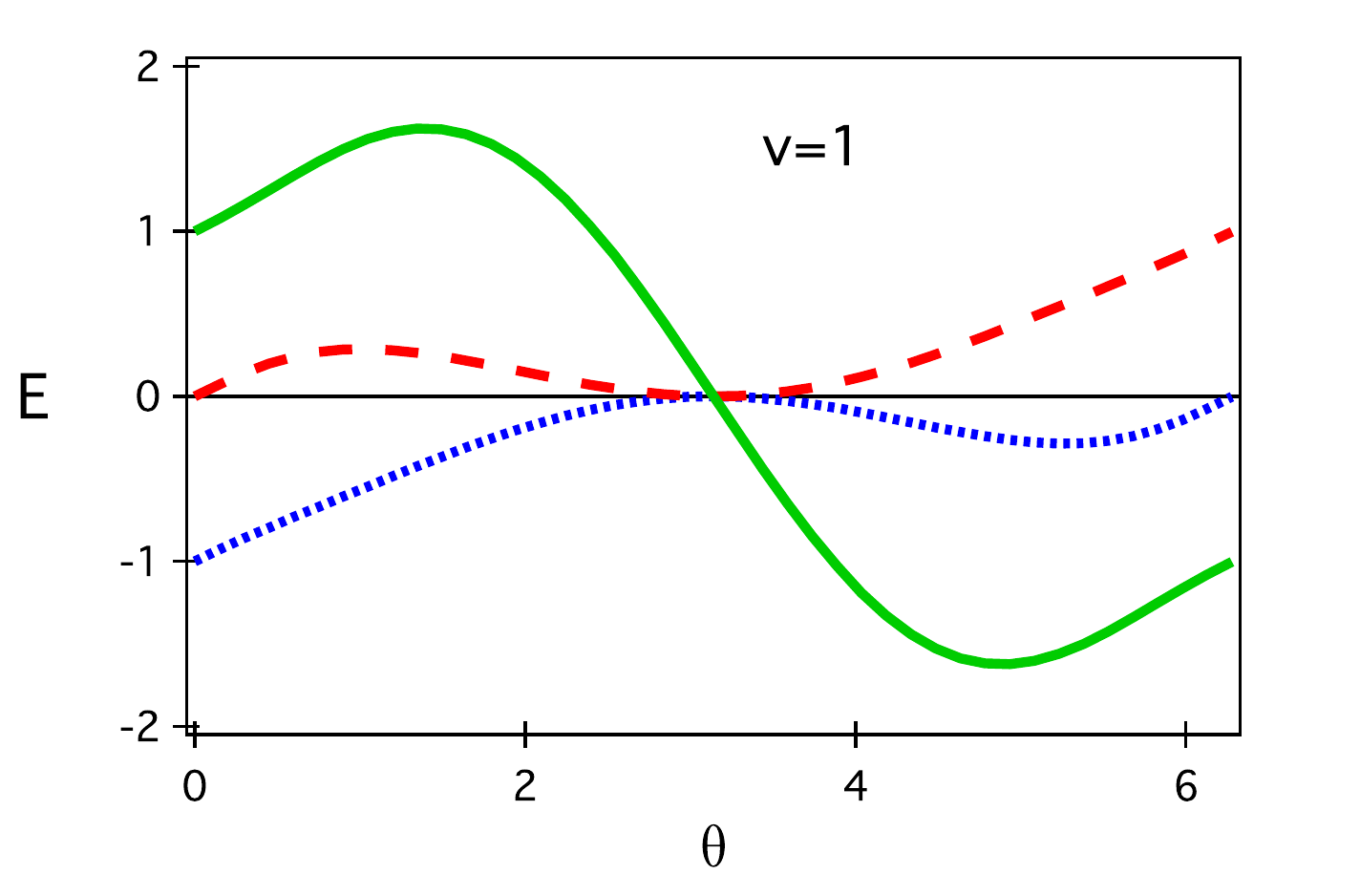}
  \caption{Energy eigenvalues $E_n(\theta)$ of the model (24) with
   $v=1$. Exotic eigenvalue holonomy is clearly observed.}
  \label{fig:m6}
\end{figure}

The eigenvalue equation
\begin{eqnarray}
H(\theta) \Psi_n(\theta) = E_n(\theta) \Psi_n(\theta)
\end{eqnarray}
is analytically solvable with eigenvalues given by
\begin{eqnarray}
E_n(\theta) = R(\theta) \cos\frac{\theta}{2} Q_n(\theta),
\end{eqnarray}
and eigenstates
\begin{eqnarray}
\label{x3wf}
\Psi_n(\theta) =
 \frac{ 1 }{ \sqrt{3Q_n(\theta)^4+1} }
 \begin{pmatrix} Q_n(\theta)(Q_n(\theta)+1) \cr
               Q_n(\theta)^2-1 \cr
                Q_n(\theta)(Q_n(\theta)-1) \end{pmatrix},
\end{eqnarray}
for $n = 1, 2, 3$,  with
\begin{eqnarray}
\!\!\!\!\!\!\!\!\!\!\!\!\!\!
Q_n(\theta)
\!=\!
\frac{  {\rm sgn } \left( \cos\frac{\eta(\theta)}{2} \right) }
       { \sqrt{3(1-\sin^\frac{2}{3}\frac{\eta(\theta)}{2}}) }
 \left[ \sin^\frac{1}{3}\frac{\eta(\theta)}{2}
       -2\sin\frac{\eta(\theta)-2\eta_n}{6} \right] ,
\end{eqnarray}
in which the angle $\eta(\theta)$ is defined by
\begin{eqnarray}
\eta(\theta) =
2 \arcsin \left[ \left(
 \frac{ 3v^2 \tan^2\frac{\theta}{2} }{ 1+ 3v^2 \tan^2\frac{\theta}{2} }
             \right)^\frac{3}{2} \right] ,
\end{eqnarray}
and the state dependent shift  $\eta_n = (2n-4)\pi$.
The function  $\eta(\theta)$ is monotonously increasing and
maps $\theta \in [0,2\pi)$ to $\eta \in [0,2\pi)$.
One example of the energy eigenvalues as function of
environmental parameter $\theta$ is shown in Fig. 6.
All eigenvalues are degenerate at $\theta=\pi$,
as a consequence of vanishing Hamiltonian $H(\pi)=0$.
%
%

We note that $Q_n(\theta)$ is $6\pi$-periodic. Moreover, we have
$Q_n(\theta+2\pi) = Q_{n+1}(\theta)$, where the subscripts are to be
understood in the sense of modulo three,
which clearly signifies the existence of exotic holonomy,
\begin{eqnarray}
E_n(\theta+2\pi) = E_{n+1}(\theta) ,
\quad
(n=1,2,3).
\end{eqnarray}
%
%

%
\begin{figure}[h]
  \centering
  \includegraphics[width=6cm]{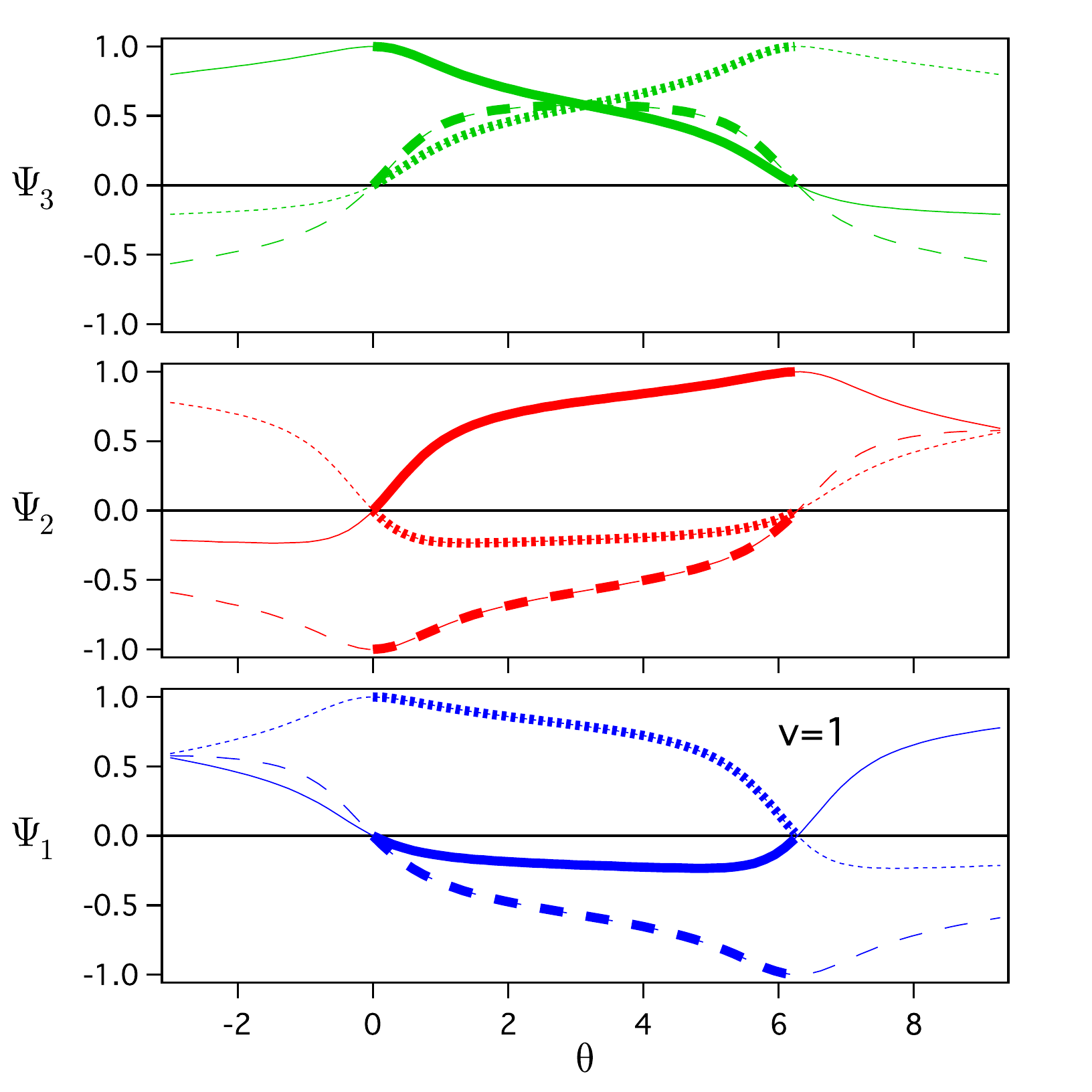}
  \caption{Eigenstates $\Psi_1(\theta)$ (bottom), $\Psi_2(\theta)$ (middle) and
   $\Psi_3(\theta)$ (top) of the model (24) with  $R(\theta)=\cos\frac{\theta}{2}$,
   and $v=1$. The solid, the dotted, and the dashed lines represent the upper, the middle, and the bottom components of the eigenvectors, respectively, all of which are chosen to be real.  See also the caption of Fig. 3.}
  \label{fig:m7}
\end{figure}
The structure of the eigenstates becomes clearer with the re-parameterization
of $Q_n(\theta)$ with a new angle variable $\xi = \xi(\theta)$ ;
\begin{eqnarray}
Q_n(\xi) =  {\rm Rt \,}\frac{\xi+\eta_n}{6} ,
\end{eqnarray}
namely, $Q_1(\xi) =  {\rm Rt \,}[(\xi-2\pi)/6]$,
$Q_2(\xi) =  {\rm Rt \,}[\xi/6]$
and $Q_3(\xi) = {\rm Rt \,}[(\xi+2\pi)/6]$,
where a $\pi$-periodic function $ {\rm Rt }$ is defined by
\begin{eqnarray}
{\rm Rt \,} \xi = \frac{1}{3^\frac{1}{4}}{\rm sgn}(\tan \xi) \sqrt{ | \tan \xi | }
 = \frac{ {\rm Rs \,} \xi  } {  {\rm Rc \,} \xi },
\end{eqnarray}
along with  $2\pi$-periodic functions $ {\rm Rs }$ and  $ {\rm Rc }$ defined as
\begin{eqnarray}
&&
{\rm Rs \,} \xi = \frac{1}{3^\frac{1}{4}}{\rm sgn}(\sin \xi) \sqrt{ | \sin \xi | },
\nonumber \\
&&
{\rm Rc \,} \xi = {\rm sgn}(\cos \xi) \sqrt{ | \cos \xi | }.
\end{eqnarray}
The functions ${\rm Rt}$, ${\rm Rs}$ and ${\rm Rc}$ are
analytic on a single sheet complex $\theta$ plane
in contrast to $\sqrt{\tan\xi}$, $\sqrt{\sin\xi}$,
and $\sqrt{\cos\xi}$, respectively,
which are analytic on a double sheet  $\theta$ plane.
The monotonously increasing function $\xi(\theta)$ maps
$\theta \in [0, 2\pi)$ to $\xi \in [0, 2\pi)$.
The eigenstates is written, with the new angle parameter $\xi$, as
\begin{eqnarray}
\label{x3wfx}
\Psi_n(\xi) =
 \begin{pmatrix}
  {\rm Rs \,}\frac{\xi+\eta_n}{6}
      \left( {\rm Rs \,}\frac{\xi+\eta_n}{6} - {\rm Rc \,}\frac{\xi+\eta_n}{6}
      \right) \cr
 {\rm Rs^2 \,}\frac{\xi+\eta_n}{6} - {\rm Rc^2 \,}\frac{\xi+\eta_n}{6}  \cr
  {\rm Rs \,}\frac{\xi+\eta_n}{6}
      \left( {\rm Rs \,}\frac{\xi+\eta_ni}{6} - {\rm Rc \,}\frac{\xi+\eta_n}{6}
      \right) \cr
   \end{pmatrix}.
\end{eqnarray}
From this from, we see that $\Psi_n(\theta)$ is {\em $6\pi$-periodic}.
The gauge potential $A_{nm}(\theta)$, which determines the
adiabatic variation of eigenstates, is given by
\begin{eqnarray}
&&
\!\!\!\!\!\!\!\!\!\!\!\!\!\!\!\!\!
A(\theta)
=
 \left[\! \begin{array}{ccc} 0 & -i & 0 \cr  i & 0 & 0 \cr 0 & 0 & 0
        \end{array}
 \! \right]  \!g(\theta+2\pi)
\\ \nonumber
&&
\!\!\!\!\!
- \left[\! \begin{array}{ccc} 0 & 0 & -i \cr  0 & 0 & 0 \cr i & 0 & 0
        \end{array}
 \! \right]  \!g(\theta)
+ \left[\! \begin{array}{ccc} 0 & 0 & 0 \cr  0 & 0 & -i \cr i & 0 & 0
        \end{array}
 \! \right]  \!g(\theta-2\pi)
,\end{eqnarray}
where $g(\theta)$ is defined by
\begin{eqnarray}
g(\theta) = \frac{1}{6} \frac{\partial \xi(\theta)}{\partial \theta}
\left< \Psi_1(\xi(\theta)) \right| i\partial_\xi \left.\! \Psi_3(\xi(\theta)) \right> .
\end{eqnarray}
The calculation of the holonomy matrix involves fully
ordered
matrix integral, thus no simple analytical calculation can be performed.
However, we can deduce from (\ref{x3wfx}), that it is given by
\begin{eqnarray}
\label{3m3}
M =
 \left[ \begin{array}{ccc}
          0 & 1 & 0 \cr  0 & 0 & 1 \cr 1 & 0 & 0
        \end{array} \right]
\label{M}
\end{eqnarray}
showing the spiral type exotic holonomy with Manini-Pistolesi phase $1$ for
all states.
\begin{figure}[h]
  \centering
  \includegraphics[width=4.2cm]{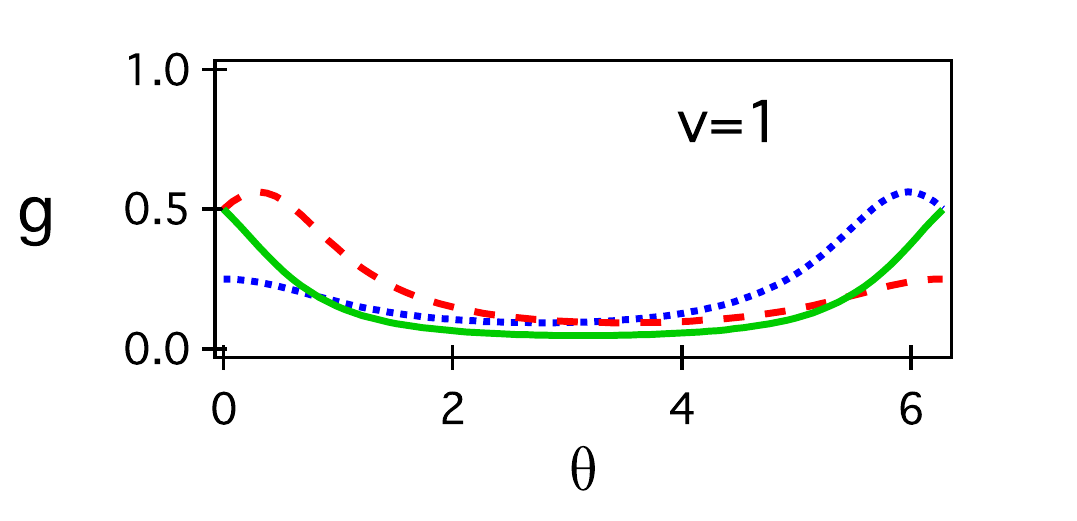}
  \caption{Functional form of gauge potential $g(\theta)$ (solid line),
  $g(\theta-2\pi)$ (dashed line) and $g(\theta+2\pi)$ (dotted line) of the model (24)
  with $v=1$.}
  \label{fig:m8}
\end{figure}

As in the two level case, we examine exceptional points $\theta^\star$ where
two complex energies coalesce, $E_m(\theta^\star)-E_n(\theta^\star)=0$.
We find
\begin{eqnarray}
\theta^\star=\pi, \quad
\theta^{\star(12)}_\pm,
\quad \theta^{\star(23)}_\pm,
\end{eqnarray}
where the first solution comes
from  $\cos\frac{\theta^\star}{2}=0$,
while $\theta^{\star(12)}_\pm$ and $\theta^{\star(23)}_\pm$
are obtained from $P_2(\theta^\star)-P_1(\theta^\star)=0$
and $P_3(\theta^\star)-P_2(\theta^\star)=0$, respectively.
We then immediately obtain
\begin{eqnarray}
&&\!\!\!\!\!\!\!\!
\theta^{\star(12)}_\pm = -2 {\rm\ arccot}
\left( \sqrt{3(e^{\pm i \frac{3}{2}\pi}-1)} v \right),
\nonumber \\
&&\!\!\!\!\!\!\!\!
\theta^{\star(23)}_\pm = 2 {\rm\ arccot}
\left( \sqrt{3(e^{\mp i \frac{3}{2}\pi}-1)} v \right).
\end{eqnarray}
Near the exceptional points, $A(\theta)$s are approximated as
\begin{eqnarray}
A_{jk}(\theta) = -A_{jk}(\theta)
\approx \mp \frac{i}{4} \frac{1}{\theta-\theta^{\star(jk)}_\pm}
\qquad
(\theta \to \theta^{\star(jk)}_\pm).
\end{eqnarray}
The existence of the poles explain the non-vanishing values
of $M_{12}$, $M_{23}$, and $M_{31}$,
implying the existence of exotic holonomy.
Fig. 9 shows the locations of poles and branching structure of energy
surface $E_n(\theta)$ in Mercator representation of
the complex parameter plane $\theta$.
\begin{figure}[h]
  \centering
  \includegraphics[width=3.5cm]{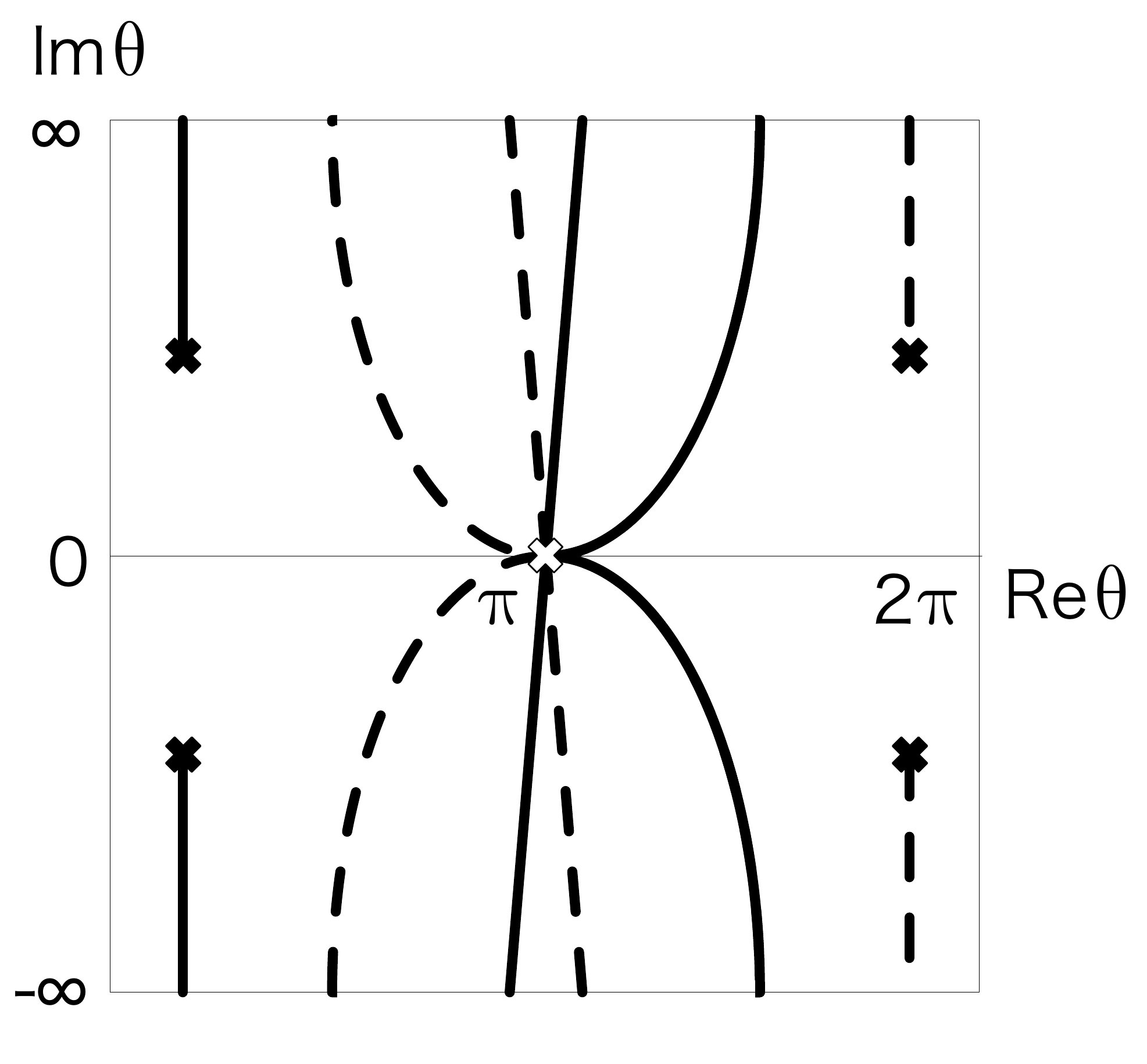}
  \caption{Exceptional points on the Mercator projection of the complex $\theta$ plane
 of system described by (24).
  The filled crosses represent the exceptional points that are the poles of gauge
  potential $A(\theta)$, while the unfilled cross is the point of eigenvalue degeneracy
  having no effect on
  the singular behavior of $A(\theta)$. The solid and the dashed lines 
  represent the branch cuts satisfying ${\rm Re}(E_3-E_2)=0$
  and ${\rm Re}(E_2-E_1)=0$, respectively.}
  \label{fig:m9}
\end{figure}

All the results obtained here do not depend on the specific choice
of the matrices (\ref{ZF3}). 
In fact, we can enlarge our model by replacing the two matrices $Z^{(3)}$ and
$F^{(3)}$ by
\begin{eqnarray}
Z^{(3)}=\Sigma_4+c_5\Sigma_5 ,
\ \
F^{(3)}=I^{(3)}+c_1\Sigma_1+c_2\Sigma_2+c_3\Sigma_3 ,
\end{eqnarray}
where $c_j$ are real numbers, $I^{(3)}$ three-dimensional
unit matrix and $\Sigma_j$ given by
\begin{eqnarray}
&&\!\!\!\!\!\!\!\!\!\!\!\!\!\!\!\!\!\!\!\!
\Sigma_1 =  \begin{pmatrix} 0 & 0 & 0 \cr 0 & 0 & 1 \cr 0 & 1 & 0 \end{pmatrix} ,
\ \
\Sigma_2 =  \begin{pmatrix} 0 & 0 & 1 \cr 0 & 0 & 0 \cr 1 & 0 & 0 \end{pmatrix} ,
\ \
\Sigma_3 =  \begin{pmatrix} 0 & 1 & 0 \cr 1 & 0 & 0 \cr 0 & 0 & 0 \end{pmatrix} ,
\nonumber \\
&&\!\!\!\!\!\!\!\!\!\!\!\!\!\!\!\!\!\!\!\!
\Sigma_4 =  \begin{pmatrix} 1 & 0 & 0 \cr 0 & 0 & 0 \cr 0 & 0 & -1 \end{pmatrix} ,
\ \
\Sigma_5 =  \frac{1}{\sqrt{3}}
  \begin{pmatrix} 1 & 0 & 0 \cr 0 & -2 & 0 \cr 0 & 0 & 1 \end{pmatrix} .
\end{eqnarray}
If we were to make independent choice of six parameters,
$R \in (-\infty, \infty)$, $\theta \in [0, 2\pi)$, $c_j \in (-\infty, \infty)$, ($j=1, 3, 4, 5$),
the Hamiltonian (\ref{3bdh}) is the most general
real symmetric three-by-three matrix.
By fixing $c_j$s and binding $R$ and $\theta$ by $R(\theta)=\cos\frac{\theta}{2}$,
we go back to the same game of
considering the system as a function of a single parameter $\theta$ which forms
a ring $S^1$.
It can be checked numerically, that
 the exotic holonomy characterized by
(\ref{3m3}), or equivalently, the {\it eigenvalue flow}
$\{1, 2, 3\} \to \{2, 3, 1\}$, in obvious notation,  is
a common characteristics of system with $c_1=c_2=c_3$.
Here, we have made an assumption that the unperturbed spectrum
is not much different from the original model, $|c_5| \ll 1$,
All possible patterns of eigenvalue flow are obtained
with suitable choice of $c_j$s.
Specifically, $c_1 \ne 0$, $c_2=c_3=0$ results in $\{1,2,3\} \to \{2, 1, 3\}$,
$c_2 \ne 0$, $c_3=c_1=0$ in $\{1, 2, 3\} \to \{3, 2, 1\}$,
and
$c_3 \ne 0$, $c_1=c_2=0$, in $\{1, 2, 3\} \to \{1, 3, 2\}$.
Generic case $c_1 \ne c_2 \ne c_3$ also produces the
second pattern,  $\{1, 2, 3\} \to \{3, 2, 1\}$.
%
It is now clear, that there is finite subset of parameter space, in which
exotic holonomies of various types arise
after cyclic variation of the parameter $\theta$.


\section{Outlook}

Our results obtained in the two and three level cases can be
extended to $N$ levels ($N\le 4$) in a straightforward way.
It is possible to prove the existence of the exotic holonomy
for systems described by the Hamiltonian
\begin{eqnarray}
\label{hngen}
H(\theta)
= \cos\frac{\theta}{2}\left[\cos\frac{\theta}{2} Z^{(N)}
+ v\sin\frac{\theta}{2} |w^{(N)}\rangle\langle w^{(N)}| \right]
,
\end{eqnarray}
where $|w^{(N)}\rangle$ is a normalized $N$-dimensional vector
and $Z^{(N)}$ is an $N\times N$ Hermitian matrix, as long as all
eigenvectors of $Z^{(N)}$ have non-zero overlap with $|w^{(N)}\rangle$. 

The Hamiltonian exotic holonomy
shares a common feature
with the Wilczek-Zee holonomy of having
 {\it SU}($N$) non-Abelian gauge potential at their base.
However, they are distinct in that, in the former,
$N$ eigenstates are exchanged among themselves with their internal dynamics, 
while, in the latter, involvement of another eigenstate, or a set of degenerate eigenstates \cite{WZ84} is required.

The exotic holonomy also seems to have resemblance to the
off-diagonal holonomy of Manini and Pistolesi.
It is important to point out that the eigenstates are obtained independently from
the choice of envelope function $R(\theta)$. If we make the choice $R(\theta)=1$, 
the Hamiltonian becomes anti-periodic, $H(\theta+2\pi)=-H(\theta)$, 
so that new period is now $4\pi$.
The eigenstate holonomy, occurring now at the midpoint of the new full cycle $\theta \in [0, 4\pi)$, 
is nothing but the off-diagonal holonomy discussed by Manini and Pistolesi.
In the off-diagonal holonomy, the set of eigenstates at the starting value of environmental
parameter ``accidentally'' coincides with that at another value of parameter
in a midpoint of cyclic evolution. In general, such coincidence is highly unlikely, and
it is often a result of the same Hamiltonian multiplied by
different numbers appearing at different value of environmental parameter.
%
In such a case,
with the introduction of new envelope function and
reinterpretation of the period of parameter variation, a system with off-diagonal holonomy 
can be mapped to another one with exotic holonomy.
In this instance, the Manini-Pistolesi holonomy {\it is} the exotic holonomy in disguise.
\begin{figure}[h]
  \centering
  \includegraphics[width=3.5cm]{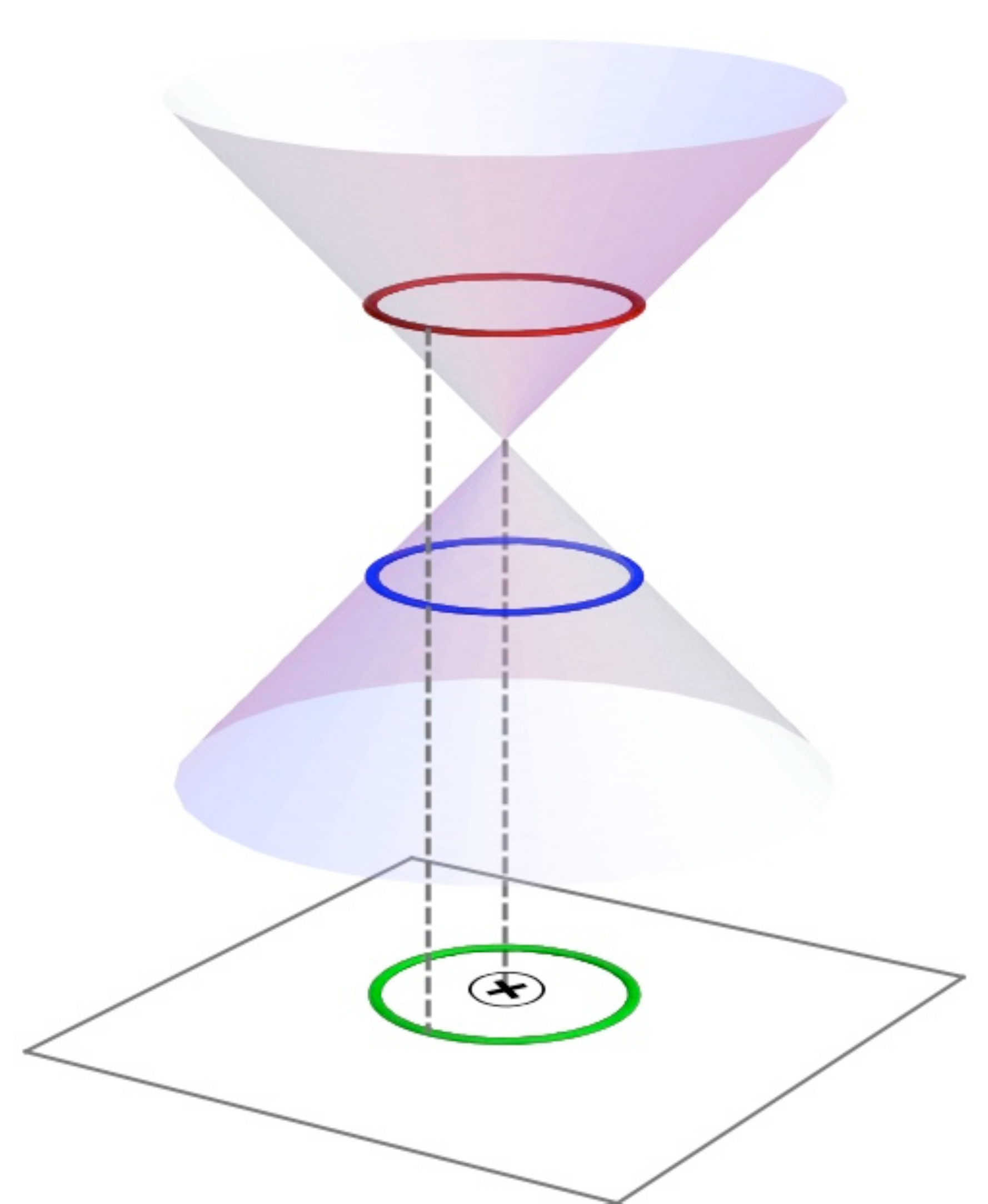}
  \includegraphics[width=3.5cm]{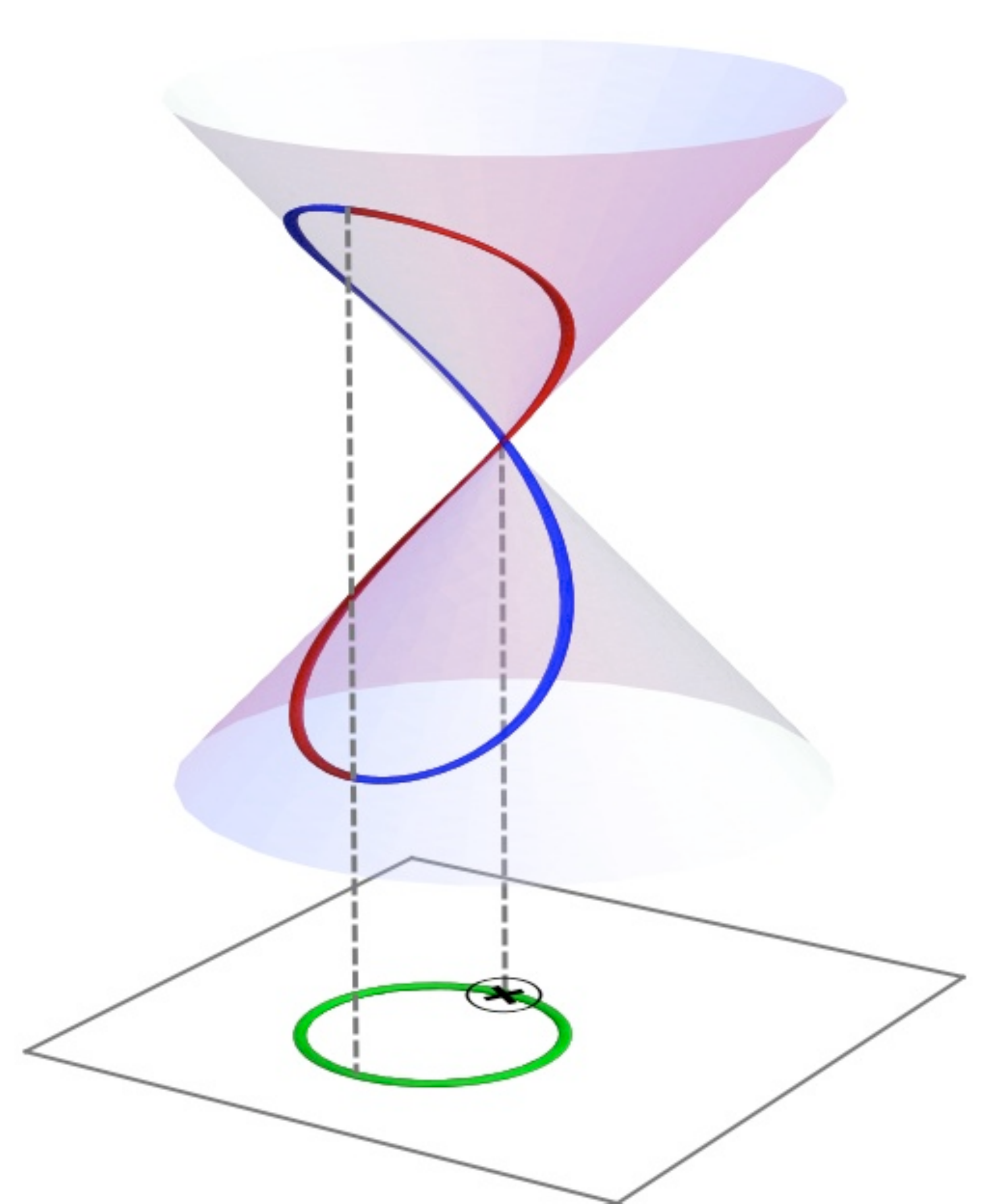}
  \caption{Two types of Hamiltonian holonomy in the presence of a diabolical point
  (cross surrounded by small circle)
  on energy surface standing on a parametric plane.  
  The picture on the left represents circular parameter variation
  that results in the Berry and Wilczek-Zee holonomies, while the picture on the right
  depicts the one leading to the exotic holonomy.
}
  \label{fig:m10}
\end{figure}

Our findings can be placed in context by considering a standard 
double-cone structure
of energy surface standing on the parameter space,
whose connected apices of two cones represent Berry's diabolical point.
%
When the parameters are varied along a circle that surrounds the diabolical point, 
Berry phase arises
(Fig.~\ref{fig:m10}, left).
We can ask a question: what will happen when the circle touches the diabolical point.
Obviously,
the trajectory on the energy surface should be ``smooth'', and it wanders both cones. 
With a cyclic variation of parameters,  the trajectory moves from one cone to the other.
This is exactly the Hamiltonian exotic holonomy (Fig. \ref{fig:m10}, right).
This process is fully described by holonomy matrix given in terms of the path-ordered integral of the gauge potential
just as in the case of Berry phase.
The gauge potential now has 
singularities in complexified parameter space, not on the diabolical point itself.
%

The Hamiltonian exotic holonomy can be viewed as an extension of, 
and a natural complement to the Berry phase, and
it forms an integral part of physics of adiabatic quantum control.
The general equation for quantum holonomy (\ref{MHOL}) is just a very natural expression of the basic requirement that the entire set of eigenstates is
to be mapped to itself after the cyclic variation of environmental parameter.
%
%

\section*{Acknowledgements}
We acknowledge the financial support by the Grant-in-Aid for Scientific Research of
Ministry of Education, Culture, Sports, Science and Technology, Japan
(Grant number 21540402),
and by  Korea Research Foundation Grant (KRF-2008-314-C00144).
%
%



\begin{thebibliography}{99}
%
\bibitem{BE84}
M. V. Berry, Proc. Roy. Soc. London {\bf A 430}, 405 (1984)
%
\bibitem{SW89}
A. Shapere and F. Wilczek, eds., {\em Geometric phases in
physics} (World Scientific, Singapore, 1989).
%
\bibitem{WZ84}
F. Wilczek and A. Zee, Phys. Rev. Lett. {\bf 52}, 2111 (1984).
%
\bibitem{ZR99}
P. Zanardi and M. Rasetti, Phys. Lett. {\bf A 264}, 94 (1999).
%
%
\bibitem{TM07}
A. Tanaka and M. Miyamoto, Phys. Rev. Lett. {\bf 98}, 160407 (2007).
%
\bibitem{MT07}
M. Miyamoto and A. Tanaka, Phys. Rev. {\bf A 76}, 042115 (2007).
%
\bibitem{CH98}
T. Cheon, Phys. Lett. {\bf A 248}, 285 (1998)
%
\bibitem{TF01}
I. Tsutsui, T. Fulop, and T. Cheon, J. Math. Phys. {\bf 42}, 5687 (2001).
%
%
\bibitem{KA50}
T. Kato,
J. Phys. Soc. Jpn. {\bf 5} (1950) 435-439.
%
\bibitem{KCT09}
S.W. Kim, T. Cheon and A. Tanaka, arXiv.org: 0902.3315 (2009).
%
 \bibitem{JS04}
N. Johansson and E. Sj{\"o}qvist,
Phys. Rev. Lett. {\bf 92} (2004) 060406.
%

\bibitem{MO07}
S.~Morita, J. Phys. Soc. Jpn. {\bf 76} (2007) 104001.

\bibitem{LA32}
L.D. Landau,
Sov. Phys. {\bf 2} (1932) 46?51.
%
 \bibitem{ZE32}
C. Zener,
Proc. Roy. Soc. London, {\bf A 137} (1932) 692--702.
%
\bibitem{CT09}
T. Cheon and A. Tanaka,
Europhys. Lett. {\bf 85} (2009) 20001(5p).
%
 \bibitem{TC09}
A. Tanaka and T. Cheon,
Ann. of Phys. (NY) {\bf 324} (2009) 1340-1359.
%
 \bibitem{MP00}
N. Manini and F. Pistolesi,
Phys. Rev. Lett. {\bf 85} (2000) 3067-3070.
%
 \bibitem{FS05}
S. Fillip and E. Sj{\"o}qvist,
Phys. Lett. A {\bf 342} (2005) 205-212.
%
%

%
%
%
%
%
%
%

\end{thebibliography}
\end{document}